# Analysis of retinal and choroidal images measured by laser Doppler holography

L. Puyo*[a], M. Paques[b,c], M. Fink [a], J-A Sahel[b,c], M. Atlan[a]
[a] Institut Langevin, CNRS, PSL Research University, ESPCI Paris, 1 rue Jussieu. Paris. France
[b] Institut de la Vision, 17 rue Moreau, 75012 Paris. France
[c] Centre d'investigation clinique des Quinze-Vingts. INSERM. 28 rue de Charenton, Paris. France

## ABSTRACT

Laser Doppler holography (LDH) is a full-field imaging technique that was recently used in the human eye to reveal blood flow contrasts in the retinal and choroidal vasculature non-invasively, and with high temporal resolution. We here demonstrate that the ability of LDH to perform quantitative flow measurements with high temporal resolution enables arteriovenous differentiation in the retina and choroid. In the retina, arteries and veins can be differentiated on the basis of their respective power Doppler waveforms. Choroidal arteries and veins can instead be discriminated by computing low and high frequency power Doppler images to reveal low and high blood flow images, respectively.

## 1. INTRODUCTION

The study of the retinal and choroidal circulations is of major interest to improve our understanding of increasingly prevalent diseases such as diabetic retinopathy, age-related macular degeneration (AMD), glaucoma, and hypertension. In the last decade, the irruption and development of optical coherence tomography (OCT) and subsequently of OCT-angiography (OCT-A) has allowed for a comprehensive study of the fundus vasculature anatomy in health and disease [1, 2], but these technologies have so far not been able to demonstrate blood flow dynamics measurements due to their limited temporal resolution. Other techniques such as laser speckle flowgraphy are able to evidence retinal hypoperfusion inherent to some ocular diseases as is the case in glaucoma [3]. Laser Doppler flowmetry (LDF) is another blood flow monitoring technique based on a short-time Fourier transform method, carried out to analyze the time profile characteristics of the Doppler spectrum to extract blood flow contrast with a temporal resolution sufficient to observe the changes transient to cardiac cycles [4]. However, this technique is performed only at a single location, and scanning implementations with the aim of imaging a two dimensional (2-D) field of view have to make a tradeoff between temporal resolution and the spatial extent of the investigated area. Thus so far flying spot and line-scanning LDF have not demonstrated hemodynamics measurements over a 2-D field of view [5, 6].

Laser Doppler holography (LDH) is a full-field imaging technique that was recently used in the human eye, to reveal blood flow contrasts in retinal and choroidal vascular networks, non-invasively, with high temporal resolution [7, 8]. It consists in the measurement of optical holograms acquired with near infrared radiation (785 nm) with an ultrahigh-speed camera, followed by a temporal Fourier transform analysis of the digital holograms. Several contributions such as pulsatile blood flow in retinal and choroidal capillaries and large vessels, ocular movements and optical diffusion participate to the power Doppler signal measured in LDH. We here demonstrate the use of two different processes to analyze the reconstructed power Doppler images in order to achieve an arteriovenous differentiation in the retina and choroid. In the retina, the process is based on analyzing systolodiastolic blood flow variations on the full-field of view, and as arteries present greater variations of blood flow, they can be easily discriminated from veins. This analysis is carried out through the calculation of the pixel-wise resistivity index. Then we take advantage of the fact that LDH allows to reconstruct separate images corresponding to different ranges of Doppler broadening. Two different power Doppler images corresponding to the low and high frequency shifts are built. In the central vision, these two images can each preferentially reveal a certain type of vessels.

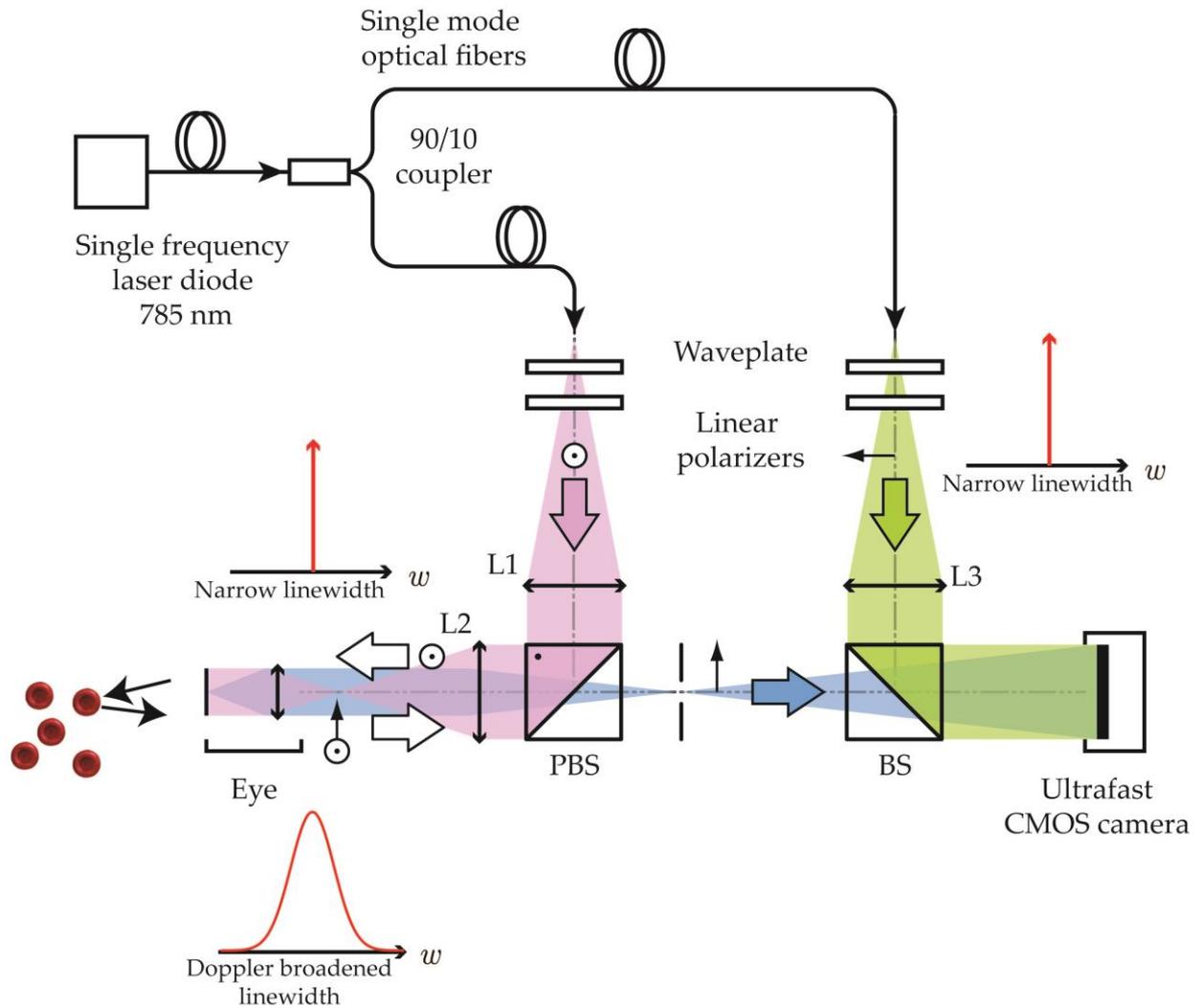

Figure 1. Optical setup. L1, L2 and L3: converging lenses. PBS: Polarizing Beam-Splitter. BS: Beam-Splitter. Light source: single frequency laser diode (SWL-7513-H-P, Newport). The retina is illuminated with the narrow frequency laser, then the Doppler broadened beam is combined with the reference beam. The resulting interference pattern is recorded with a CMOS camera (Phantom V2511, Ametek) running at a frame rate up to 75 kHz in 512x512 format.

## 2. METHODS & RESULTS

The experimental setup used for this study, shown in Fig. 1, consists of a fiber Mach-Zehnder optical interferometer. The light source is a 45 mW single-mode, fiber laser diode (Newport SWL-7513-H-P) at wavelength 785 nm, spatially and temporally coherent. The retina is illuminated with 1.5 mW of constant exposure over 4 x 4 mm. A Polarizing Beam Splitter (PBS) cube is used in the object arm to illuminate the eye and collect the light backscattered by the retina with the camera. The object and collimated reference waves are combined using a non-polarizing beam splitter cube and they interfere on the sensor plane. The polarization of the reference wave is adjusted with a half-wave plate and a polarizer to optimize fringe contrast. Interferograms are recorded on a fast CMOS camera (Ametek - Phantom V2511, 60 or 75 kHz, 512x512 format, quantum efficiency 40%, 12-bit pixel depth, pixel size 28 µm) and are processed offline using Matlab. The raw interferograms are first propagated numerically using the angular spectrum propagation method. Except in the cases of unusually swollen choroid, the depth of field of the images is large enough so that both the retinal and choroidal vasculatures are reconstructed in the same holographic image. Then in our approach, cross-beating terms of the interferograms carry the optical Doppler broadening and data processing consists of measuring the local optical temporal fluctuations. To that end a short-time Fourier transform analysis is performed with a window size of 512 images (twin) and an overlap between two consecutive windows of 256 images. The power Doppler images are obtained by integrating the highest frequencies of the Fourier transform of the squared magnitude of the reconstructed holograms. The frequency threshold is generally set between 6 and 10 kHz. Once the power Doppler movie sequence of the LDH measurement has been calculated, the intensity variations are analyzed pixel-wise. The standard deviation and average power Doppler images are calculated and their ratio yields the coefficient of variation image. Because they have greater blood flow variations during cardiac cycles than in veins, arteries come out in red. The process used to obtain the arteriovenous maps in the retina from the raw interferograms is illustrated in Fig. 2.

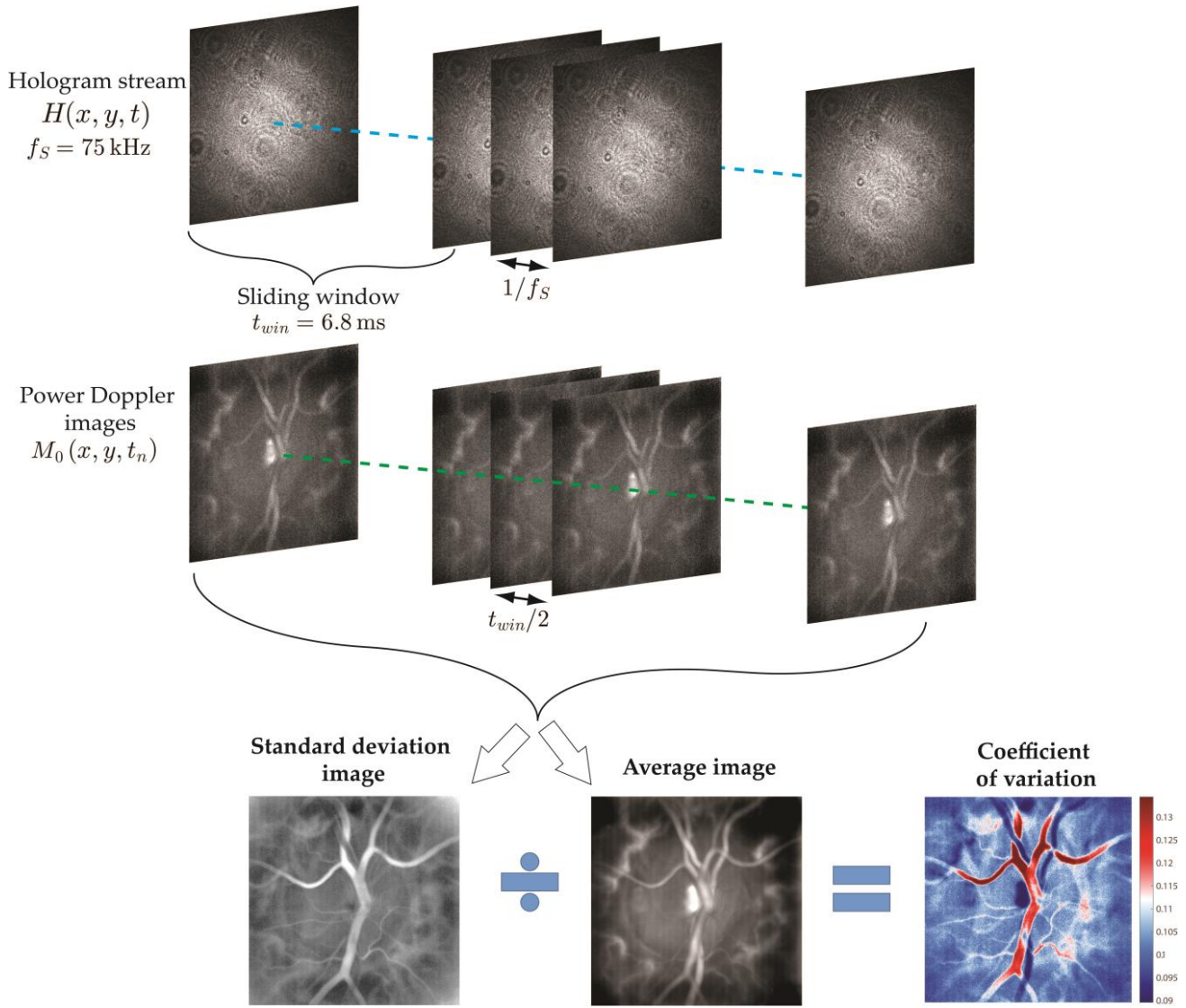

Figure 2. Processing of the holograms after numerical propagation. The short-time Fourier transform analysis is carried out to obtain a blood flow movie with a sliding window of 6.8 ms in the case of a 75 kHz measurement. The standard deviation and average power Doppler images are calculated. Their ratio yields the coefficient of variation image; with this colormap arteries come out in red, and veins in dark blue.

To discriminate these contributions, we analyzed the reconstructed power Doppler waveforms using different Doppler frequency ranges. Very low frequency shifts are dominated by global eye movements such as fixational eye movements and fundus pulsations. Low frequency shifts reveal slow ocular movements and vessels carrying slow blood flows. On the contrary high frequency shifts reveal blood vessels carrying great blood flows. In practice we use the frequency ranges 2.5 – 6 kHz and 10 – 30 kHz to reveal the low and high blood flows. These two images are then merged into a single composite color image: low blood flow are encoded in cyan and high blood flow in red [8].

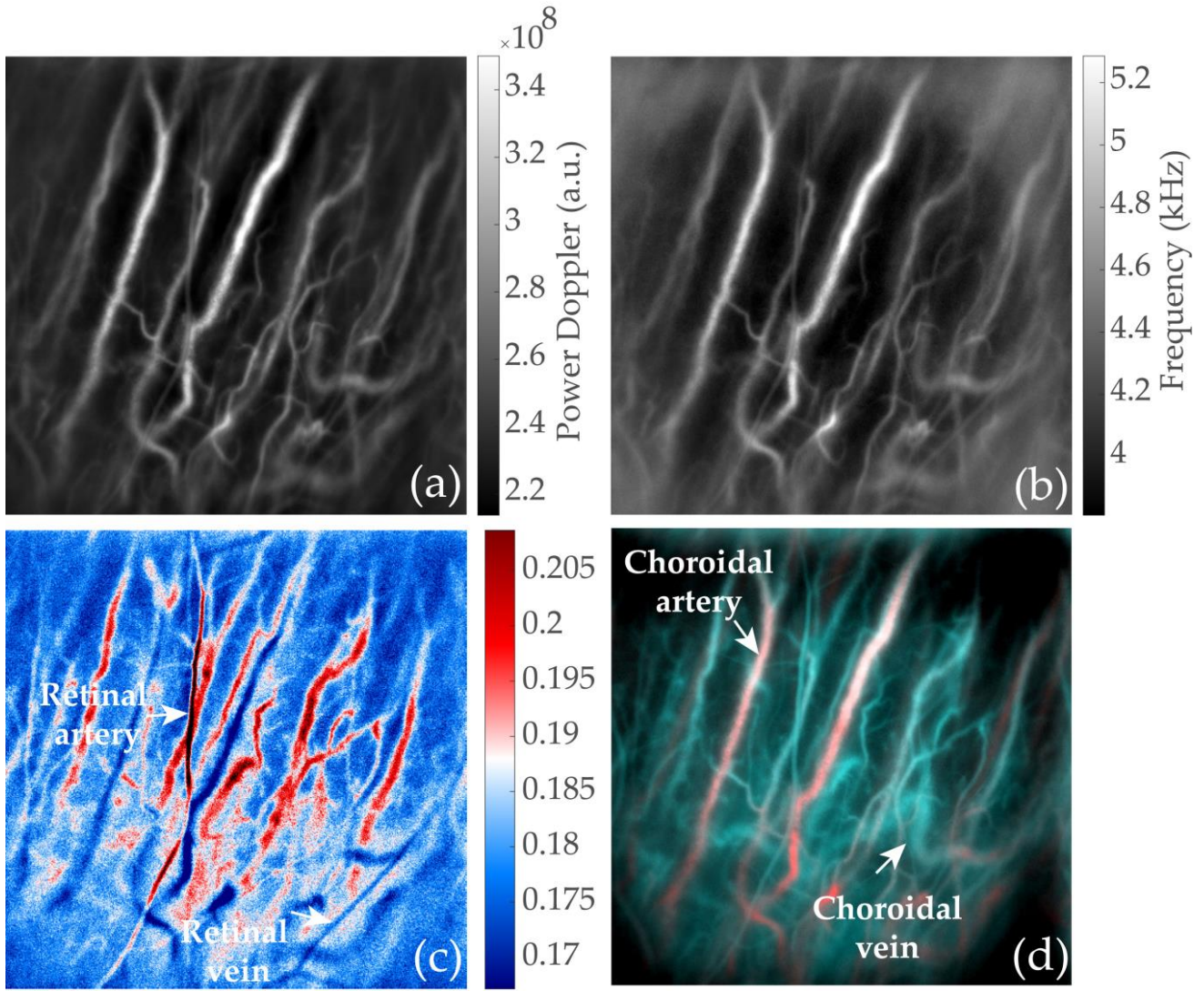

Figure 3. Different images are obtained using various image processing on the same dataset. (a) Power Doppler image showing most vessels from the retina and choroid. (b) Mean frequency shift calculated from the Doppler power spectrum density revealing vessels with a similar contrast. (c) Coefficient of variation image revealing retinal arteries in red and retinal veins in blue. (d) Composite color image revealing choroidal arteries in red and choroidal veins in blue.

In Fig. 3 we show the results of the two methods we mentioned. First in Fig. 3(a) and (b) are displayed the power Doppler and mean frequency shift images that reveal the retinal and choroidal vasculatures. The arteries and veins from both vasculature can however not be distinguished on the basis of the flow values as they are averaged over several cardiac cycles. However in the coefficient of variation map in Fig. 3(c), a difference of contrast between retinal arteries and veins can be observed. The retinal arteries come out in red due to their larger systolodiastolic variations whereas the retinal arteries come out in blue. The effect of this processing for choroidal vessels is less efficient. Finally, in the composite color image in Fig. 3(d), choroidal arteries which are visible in the high Doppler frequency shift image (not displayed here) come out in red, whereas choroidal veins come out in blue because they are revealed in the low frequency shift image. LDH is thus able to distinguish arteries and veins from both the retinal and choroidal layers.

## 3. CONCLUSION

We have found that wideband laser Doppler holography operating at 785 nm can image the retinal and choroidal vasculatures non-invasively while keeping the retinal full-field exposure under 10 mW/cm². Power Doppler measurements at high temporal resolution in the retina allow to reveal the specific blood flow behavior in arteries and veins. Coefficient of variations maps can be calculated from pixel-wise flow variations and lead to a retinal arteriovenous discrimination. In the choroid, the instrument is especially sensitive to large blood flows and thus preferentially reveals choroidal arteries and arterioles, even at large depth. Additionally, bandpass filtering the Doppler spectrum allows to separate vessels according to their flows which opens possibilities for basic flow analysis in the choroid. Indeed, because there exists large differences of blood flow in the choroid between arteries and veins, a basic flow analysis leads to an arteriovenous segregation.

## ACKNOWLEDGEMENTS

This work was supported by LABEX WIFI (Laboratory of Excellence ANR-10-LABX-24) within the French Program Investments for the Future under Reference ANR-10-IDEX-0001-02 PSL, and the European Research Council (ERC Synergy HELMHOLTZ, grant agreement #610110)